\documentclass{naturedummy}
\usepackage{amsmath}
\usepackage{pdfsync}
\usepackage{graphicx}

\advance \topmargin by -\headheight
\advance \topmargin by -\headsep

\evensidemargin \oddsidemargin
\newcommand{\abs}[1]{\left\vert#1\right\vert}
\newcommand{\ket}[1]{\left\vert#1\right\rangle}

\newcommand{\expect}[1]{\left\langle#1\right\rangle}

\newcommand{\up}{\left\vert \uparrow\right\rangle}
\newcommand{\down}{\left\vert \downarrow\right\rangle}

\title{Single Ion Quantum Lock-In Amplifier}

\author{Shlomi Kotler$^{1}$, Nitzan Akerman$^1$, Yinnon Glickman$^1$, Anna Keselman$^1$, \& Roee Ozeri$^1$}

\begin{document}

\maketitle

\begin{affiliations}
 \item Department of Physics of Complex Systems, Weizmann Institute of Science, Rehovot 76100, Israel
\end{affiliations}

\begin{abstract}
Quantum metrology applies tools from quantum information science to improve on measurement signal-to-noise ratio\cite{Rosenband02008,Knunz2010}. The challenge is to increase sensitivity while reducing susceptibility to noise; tasks which are often in conflict. Invented by Dicke, the lock-in measurement is a detection scheme which overcomes this difficulty by spectrally separating signal from noise. Here we report on the implementation of a quantum analog to the classical lock-in amplifier. All the lock-in operations: modulation, detection and mixing, are performed via the application of non-commuting quantum operators on the electronic spin state of a single-trapped $Sr^+$ ion. We significantly increase its sensitivity to external fields while extending phase coherence by three orders of magnitude, to more than one second. With this technique we measure magnetic fields with sensitivity of $25\ pT/\sqrt{Hz}$, and light shifts with an uncertainty below $140\ mHz$ after $1320$ seconds of averaging. These sensitivities are limited by quantum projection noise and, to our knowledge, are more than two orders of magnitude better than with other single-spin probe technologies\cite{Degen2008}. In fact, our reported sensitivity is sufficient for the measurement of parity non-conservation\cite{Fortson1993}, as well as the detection of the magnetic field of a single electronic-spin one micrometer from an ion-detector with nanometer resolution. As a first application we perform light shift spectroscopy of a narrow optical quadruple transition. Finally, we emphasize that the quantum lock-in technique is generic and can potentially enhance the sensitivity of any quantum sensor.
%During measurements, phase coherence of our quantum probe is maintained for more than one second.
\end{abstract}

%Introduction about trapped ions in the context of metrology and precision
Quantum probes are advancing the field of metrology with unprecedented sensitivities. In particular, cold trapped ions are well isolated from their environment, their internal states and motion can be controlled with high fidelity, thus enabling researchers to use them as excellent probes. Examples include ion based atomic clocks\cite{Rosenband02008}  with a fractional frequency uncertainty of $5.2 \cdot 10^{-17}$ and force measurements\cite{Knunz2010} with a sensitivity of $5\cdot 10^{-24}\ N$.

%The affect of noise on a quantum probe is generally manifested via its loss of coherence.
To achieve measurements with a high signal-to-noise ratio, one has to decrease the effect of noise on the probe while enhancing its response to the measured signal. These two tasks are in many cases contradicting since noise and signal often couple to the probe via the same physical channel. Quantum metrology science uses methods from quantum coherent control to address this difficulty. As an example, entangled states which are invariant under certain noise mechanisms were engineered with trapped-ions and have demonstrated long coherence times \cite{Kielpinski2001,Roos2004,Langer2005}. Other entangled states were similarly engineered to enhance trapped ions measurement sensitivity \cite{Leibfried2004,Roos2006}. Whether or not the measurement signal-to-noise ratio indeed improves depends on the commutativity of noise and signal operators as well as on noise bandwidth\cite{Cirac1997,Lukin2004}.

A different approach to noise reduction is based on spectrally separating a quantum system from its noise environment. Such time dynamical noise decoupling was demonstrated using trapped ion qubits, among other systems, and has been optimized to match different noise profiles\cite{Uys2009,Biercuk2009}. In fact, it was theoretically shown that the decoherence rate of these modulated systems can be used to extract information about their noise spectrum\cite{goren:2007}. A natural extension to spectra characterization is the measurement of oscillating signals. Dynamical manipulation can therefore be used to decouple a quantum probe from noise while enhancing its sensitivity to alternating signals.

In the last few years dynamical decoupling methods have been used to improve on the signal-to-noise ratio of ac-magnetometry using Nitrogen-Vacancy (NV) centers\cite{Maze2008,Hall2010,Naydenov2010,deLange2010_2}. Indeed, significant enhancement of sensitivity was achieved using few tens of modulation pulses\cite{deLange2010_2}. However, due to the particular decoherence mechanism in NV-centers, their best reported magnetic field measurement sensitivity of $4\ nT/\sqrt{Hz}$ was achieved using a single echo pulse\cite{Balasubramanian2009}.

%overview of this work
In this work we show that a quantum probe, time evolving under non-commuting operators of noise, signal and modulation, is equivalent to a lock-in amplifier. We realize a quantum lock-in amplifier using a single trapped $Sr^+$ ion. By applying hundreds of modulation pulses we reach sensitivities which, to our knowledge, are more than two orders of magnitude better than with other single-spin probe technologies\cite{Degen2008}. Specifically using trapped ions, this method is likely to have strong implications on precision measurements and metrology. As a first application we perform precision light shift spectroscopy of a narrow optical quadruple transition.

%Classical lock-in
Classical lock-in amplifiers are detectors that can extract a signal with a known carrier frequency from an extremely noisy environment. Schematically, if noise $N(t)$ adds to a physical observable, $S_0$, oscillating at a frequency $f_m$, the total signal measured by the detector is $M(t)=m_0[S_0\cos(2\pi f_m t+\varphi)+N(t)]$. Here, $m_0$ sets the detector measurement units and $\varphi$ is a constant phase. A signal proportional to $S_0$ is obtained by a mix-down process: $M(t)$ is multiplied by either $\sin(2\pi f_m t)$ or $\cos(2\pi f_m t)$ and the two results are integrated over an integration window $T$,
\begin{equation}  \label{eq:classicalLI}
\begin{split}
I_{lock-in}&=\frac{1}{T}\int_0^T M(t)\cdot \cos(2\pi f_m t) dt\\
Q_{lock-in}&=\frac{1}{T}\int_0^T M(t)\cdot \sin(2\pi f_m t) dt.
\end{split}
\end{equation}
The signal $S_0$ is proportional to $(I_{lock-in}^2+Q_{lock-in}^2)^{1/2}$. The constant phase $\varphi$ can be extracted by $\tan(\varphi)=Q_{lock-in}/I_{lock-in}$. Noise spectral components with frequencies far from $f_m$ will be averaged out in the integration. Therefore, by choosing $f_m$ outside the noise band-width, the measurement signal-to-noise ratio can be significantly improved.

%Quantum lock-in
The main obstacle in realizing quantum lock-in dynamics is finding a quantum analog to signal multiplication, which is essential for the mix-down process. In a classical apparatus this is achieved by a non-linear device with an output that is proportional to the instantaneous product of its inputs. Non-linear dynamics of the wavefunction cannot be introduced directly due to the linearity of Schr\"odinger's equation. Nevertheless, wavefunction dynamics will be proportional to a product of Hamiltonian terms if the total Hamiltonian does not commute with itself at different times. Operator non-commutativity therefore plays an important role in the quantum mix-down process.

To show this in more detail, we turn to the case of a two-level quantum probe, with states $\up$ and $\down$. Without loss of generality, we assume that the probe is coupled both to a signal $S(t)$ and noise $N(t)$, via $H_{int}=\frac{1}{2}M(t)\hat{\sigma}_z$, where $M(t)=S(t)+N(t)$. For a lock-in measurement $S(t)$ is modulated; $S(t)=S_0\cos(2\pi f_m t+\varphi)$. The probe is initialized to $\ket{\psi_0}=\frac{1}{\sqrt{2}}(\up+\down)$. In a Bloch sphere picture this state is represented by a vector along the $x$ axis. Under $H_{int}$ the superposition phase (angle between the Bloch vector and the $x$ axis) is oscillating back and forth due to $S(t)$ and is randomly varying due to the effect of noise. To implement a lock-in measurement we mix the probe phase with an oscillating signal by adding to $H_{int}$ an oscillating term which is non-commuting with $\hat{\sigma}_z$, $H=\frac{1}{2}(M(t)\hat{\sigma}_z + \Omega(t)\hat{\sigma}_y)$. If $\Omega(t)$ is periodic and synchronized with $S(t)$ then the phase accumulated due to $S(t)$ coherently adds up whereas the random phase accumulated due to $N(t)$ is averaged away. The probe superposition is characterized by the probability of finding the probe in the $\up$ state, $P_{\uparrow}$, and the superposition relative phase $\phi_{lock-in}$. By measuring both at time $T$, we extract the quantum lock-in signal\footnote{Equation (\ref{eq:accPhase}) is valid when $\abs{M(t)}/h\ll f_m$ (Weak Coupling Regime). For a more general discussion, see supplementary material.},
\begin{equation}  \label{eq:accPhase}
\begin{split}
\phi_{lock-in}&=\frac{1}{\hbar}\int_0^T dt M(t) \cos \left( \frac{1}{\hbar}\int_0^t dt' \Omega(t') \right)\\
1-2P_{\uparrow}&=\frac{1}{\hbar}\int_0^T dt M(t) \sin \left( \frac{1}{\hbar}\int_0^t dt' \Omega(t') \right).
\end{split}
\end{equation}
Equation (\ref{eq:accPhase}) resembles the classical lock-in output in Eq. (\ref{eq:classicalLI}). Specifically, for a constant $\Omega(t)\equiv \Omega_0$, the lock-in outputs $\phi_{lock-in}$ and $(1-2P_{\uparrow})$ faithfully represent the two signal quadratures. Here, instead of reading out a classical parameter, the quantum lock-in readout requires repetitive quantum projection measurements. Notice that the two signal components can be interchanged via single qubit rotations.

% Validity of the formula
%Equation (\ref{eq:accPhase}) is a useful approximation of a more general formula (see supplementary information). The main condition for its validity is that both noise and lock-in signal change the quantum probe energy states by a frequency which is much smaller than the modulation frequency, i.e. $\abs{M(t)}/h\ll f_m$. In a lock-in experiment, the modulation and mix-down signals are synchronized, i.e. $\varphi=0$, in which case the equation is valid with no further restrictions. If for some reason, the signals are not synchronized ($\varphi\neq 0$) a second condition is needed; the total phase evolution of $\phi_{lock-in}$ should be significantly smaller than $\pi$. In practice, eq. (\ref{eq:accPhase}) is a good approximation even when $\phi_{lock-in}$ is in the order of $\pi/4$. These conditions have no analogs in the classical case. Finally, equation (\ref{eq:accPhase}) is accurate without any restrictions, in the limit of ideal $\pi$-pulses, i.e. when $\Omega(t)$ is an ideal train of arbitrarily spaced infinitely short $\pi$-pulses.

%Description of our setup
In our experiment, we use the two spin states of the electronic ground level of a single $^{88}Sr^+$ ion, $\up=\ket{5s_{1/2}, m=\frac{1}{2}}$ and $\down=\ket{5s_{1/2}, m=-\frac{1}{2}}$ as a two-level quantum probe . The ion is trapped in a linear rf Paul trap, with an axial (radial) trap frequency of  $1\ MHz$ ($2.5\ MHz$), and laser-cooled to $\approx 1\ mK$. At this temperature, the ion wavefunction extent is $48\ nm$. If used, ground-state cooling would result in an extent of $9\ nm$. A magnetic field of $204\ \mu T$ sets the spin quantization axis and lifts the degeneracy between the probe states by $f_0=5.72$ MHz.  Spin rotations are performed by pulsing an oscillating magnetic field, perpendicular to the quantization axis, at $f_0$, resulting a Rabi angular frequency, $\Omega_R = (2\pi)65.8\ kHz$. The angle and direction of rotation are determined by the pulse duration and the oscillating field phase, $\phi_{rf}$, respectively. The probe state is measured by shelving the $\up$ state to the appropriate metastable $D$ sub-level, followed by state selective fluorescence\cite{Kesselman2010}. State preparation to $\up$ is done by optical pumping (see Fig. 1(a) for a level diagram). Since the probe states are first order sensitive to magnetic fields, the main noise mechanism is magnetic field noise, with dominant spectral contributions at the line $50\ Hz$ and its harmonics. Examples for signals we can measure are modulated magnetic or light fields via their Zeeman or light shifts correspondingly.

% Lock-In pulse sequence
The lock-in sequence is depicted in Fig. 1(b). Following optical pumping, a $\pi/2$ rotation initializes the ion probe to $\ket{\psi_0}=\frac{1}{\sqrt{2}}(\up+\down)$. To modulate the ion-probe we apply a train of $N$, $\pi$-pulses, $\tau_{arm}$ apart, also known as the Carr-Purcell-Meiboom-Gill pulse sequence\cite{GillMeiboom1958}. Using this sequence the cosine term in (\ref{eq:accPhase}) is a square waveform, with a $2\tau_{arm}$ period and the sine term vanishes. Therefore, a measured signal has to be modulated at $f_m=1/2\tau_{arm}$ and in phase with the ion modulation, i.e. $\varphi=0$. Here, $\phi_{lock-in}$ is proportional to the signal magnitude, $S_0$. To measure the probe phase we complete the sequence with an additional $\pi/2$ rotation, with a relative $\phi_{rf}$ phase with respect to the initial $\pi/2$ pulse. We then detect the probability of the ion to be in the $\up$ state, $P_{\uparrow}=\frac{1}{2}+\frac{A}{2}\cos(\phi_{lock-in}+\phi_{rf})$. By scanning $\phi_{rf}$ we are able to retrieve both $\phi_{lock-in}$ and the cosine fringe contrast $A$.

%phase sensitivity
The lock-in signal $\phi_{lock-in}$ is related to a frequency shift by $\delta f=\phi_{lock-in}/2\pi T$, where $T=(N+1)\tau_{arm}$ is the total duration of the lock-in sequence. The frequency shift measurement sensitivity, $s$, is\footnote{see supplementary information}:
\begin{equation}
s=\frac{1}{2\pi}\sqrt{\frac{2(4-A^2)}{A^2T}}\frac{Hz}{\sqrt{Hz}}. \label{eq:phSensitivity}
\end{equation}
Clearly, in order to optimize the measurement sensitivity, $A$ has to be maximized, where the standard quantum limit to the sensitivity is reached when $A=1$.

%noise floor 1: 50,100,150Hz noise
We initially quantify the noise-floor of our lock-in detector at different modulation frequencies, $f_m$, and lock-in sequence duration, $T$, in the absence of any modulated signal. To begin with, we perform this measurement at slow lock-in modulation frequencies, which are comparable to typical magnetic noise frequencies in our lab. We measure $A$ for $\pi$-pulse inter spacing $\tau_{arm}$ ranging from zero to $12ms$, and for $N=1$ to $17$ $\pi$-pulses per lock-in sequence. Both the lock-in sensitivity and spectral resolution increase as $N$ increases. As shown in Fig. 1(c), dips in the fringe contrast emerge as we increase $N$. These dips, marked by gray shaded stripes, correspond to ac magnetic field noise components at frequencies of $200$, $100$ and $50\ Hz$. Fig. 1(d) shows two phase scans for a $N=17$ lock-in sequence. One scan is at $\tau_{arm}=3.6\ ms$, where no noise is present and the other for $\tau_{arm}=5\ ms$, where the lock-in modulation is has the same period as the $100\ Hz$ noise component. The two scans show a $\pi$ phase shift between them.

% the fit to Bessel functions
In order to use the lock-in method to quantify the magnetic noise spectrum we assume it is composed mainly of discrete components, $f_n=\omega_n/(2\pi)$, with corresponding amplitudes $B_n$. We can therefore write\cite{preprint},
\begin{equation} \label{eq:bessel}
A(N,\tau_{arm})=\prod_{n} J_0\left(\frac{-4g\mu_B}{\hbar} \frac{B_n}{\omega_n}\sin^2\left(\frac{\omega_n\tau_{arm}}{2}\right)\frac{\sin(N\omega_n\tau_{arm})}{\sin(\omega_n\tau_{arm})}\right).
\end{equation}
Here $J_0$ is the zeroth Bessel function of the first kind, $g$ is the Land\'e g-factor and $\mu_B$ is Bohr magneton. Notice that $J_0$ can indeed have negative values which result in a $\pi$ phase shift in the lock-in phase scan. Fig. 2(a). shows Eq. (\ref{eq:bessel}) (solid line) and our measured $A$ (filled circles) for a lock-in sequence with $N=17$. Here we assume four discrete magnetic-noise spectral components, $50$, $100$, $150$ and a slowly varying field. The noise amplitudes are taken as fit parameters, yielding  $B_{50 Hz}=540(3)\ pT,\ B_{100 Hz}=390(5)\ pT,\ B_{150 Hz}=260(4)\ pT$ and $g\mu_{B}B_\textrm{slow}f_\textrm{slow}/h=37(4)Hz^2$. The relatively low magnetic field amplitudes are due to an active magnetic field noise cancelation system\cite{preprint}.

%High modulation frequencies
Observing that noise amplitudes above $200\ Hz$ are negligible, we turn to higher modulation frequencies, in search of the longest attainable probe coherence time. We modulate the ion-probe at $f_m=312.5\ Hz$ ($\tau_{arm}=1.6\ msec$). Figure 2(b) shows the measured lock-in phase scan contrast vs. $N$, up to $N=700$. Here, due to the large number of $\pi$ pulses, $\phi_{rf}$ alternates by $\pi/2$ between consecutive pulses (X-Y modulation), to prevent rotation errors from coherently accumulating. A fit to an exponential decrease in phase contrast yields a probe coherence time of $1.4(2)\ s$. This is three orders of magnitude longer than the coherence time in the absence of lock-in modulation, measured using Ramsey spectroscopy.

% sensitivity results
From the data presented thus far we can report our probe best sensitivity. We calculate the lock-in sensitivity vs. $T$, the total lock-in sequence duration, from the fitted phase scan contrast, $A$, using Eq. (\ref{eq:phSensitivity}). Figure 2(c) shows the lock-in sensitivity at the low modulation frequency range. A minimum of $1.53\ \frac{Hz}{\sqrt{Hz}}$ ($53\ \frac{pT}{\sqrt{Hz}}$) is observed at a total experiment time of $T=120\ ms$, in between noise components. Figure 2(d) shows the lock-in sensitivity vs. $T$, for $f_m=312.5\ Hz$. Here a best sensitivity of $0.72\ \frac{Hz}{\sqrt{Hz}}$ ($25.5\ \frac{pT}{\sqrt{Hz}}$) is observed at a total experiment time of $T=560\ ms$. This is, to our knowledge\cite{Degen2008}, the best magnetic field sensitivity reported so far using a single spin (or pseudo-spin) detector. In both cases the measured sensitivity is within a factor of $1.5$ from the standard quantum limit, shown by the dashed line.

% Light shift lock-in measurement.
We next demonstrate a lock-in detection of a small signal and experimentally verify Eq. (\ref{eq:accPhase}). To this end we use the light-shift of a narrow line-width ($<100\ Hz$) laser, nearly resonant with the $ \up \rightarrow |4d_{5/2}, m=3/2\rangle$ quadruple transition at $674\ nm$. The laser amplitude is switched on and off at a rate $f_L = 500\ Hz$. With this modulation scheme, both the lock-in and the signal are square-wave modulated. We apply an, $N=99$ $\pi$-pulses, lock-in sequence and scan the lock-in modulation frequency. Here the $674nm$ laser is detuned by $\delta=-17kHz$ from resonance. The laser Rabi frequency, $\Omega_0=(2\pi)840\ Hz$, is independently measured by an on-resonance Rabi nutation curve. Figure 3(a) shows a lock-in phase scan at a lock-in modulation frequency $f_m = 500\ Hz$. The solid line is a best fit to $P_{\uparrow}=\frac{1}{2}+\frac{A}{2}\cos(\phi_{lock-in}+\phi_{rf})$, with $A$ and $\phi_{lock-in}$ as fit parameters. A clear phase shift of $0.99\pi\ rad$ is observed. The columns in Fig. 3(b) are a phase scans similar to Fig. 3(a) taken at different lock-in modulation frequencies. As seen, the lock-in signal is maximal when the lock-in modulation approaches $500\ Hz$ ($\tau_{arm}=1000\ \mu s$), i.e. the modulation rate of the laser. Figure 3(c) is showing a good agreement between the fitted $\phi_{lock-in}$, for different modulation rates, (filled circles) and the prediction of Eq. (\ref{eq:accPhase}) (solid line), with no fit parameters. A light shift of $9.7(4)\ Hz$ is measured as compared to the theoretically predicted value of $9.9(4)\ Hz$.

%%Allan Variance of method & measuring single spin B field
Any measurement uncertainty is ultimately limited, at long integration times, by systematic drifts. We measure the limit to our light shift measurement uncertainty by performing a $3.5$ hours long measurement, during which $100$ consecutive lock-in phase scans are taken, and perform an Allan deviation analysis\cite{AVAR}. A minimum to the overlapping Allan deviation, $\sigma_y(\tau)=138(6)\  mHz$ is obtained at an integration time of $\tau=1320\ s$, and is most likely limited by slow frequency drifts of the $674\ nm$ laser. When removing a linear slope from the data, the minimal Allan deviation reduces to $14(5)\  mHz$ at $\tau=4320\ s$. The magnetic field generated by the valence electron spin of a single $^{88}Sr^+$ ion will cause a level shift of $52\ mHz$ to a probe-ion trapped one micrometer away, the measurement of which could be within our experimental reach.

% S->D light shift spectroscopy
Finally, we show how the lock-in method can be used to perform light shift spectroscopy. We probe the $ \up \rightarrow |4d_{5/2}, m=3/2\rangle$ transition. Fig. 4(a) shows lock-in phase scans (columns) for different laser detunings. Figures 4(b) and 4(c) show the fitted $A$ and $\phi_{lock-in}$, respectively. Population transfer to the $|4d_{5/2}, m=3/2\rangle$ level results in a reduction in $A$ whenever the laser is close to resonance. The measured light shift is seen to be dispersive around resonance. The three resonances; a carrier and two sidebands; are due to the fast amplitude modulation of the laser, reminiscent of the Pound-Drever-Hall signal of a laser scanning across an optical cavity resonance\cite{PDD}. Such dispersive signal can be used to lock a narrow line width laser to an an atomic clock transition. Here, a possible advantage is that fast laser frequency noise components are rejected due to the lock-in measurement scheme.

%Optimal pulses
We remark that although here we used a train of evenly spaced $\pi$-pulses as our modulation scheme for quantum lock-in detection we also investigated the use of the unevenly spaced $\pi$-pulses Uhrig scheme\cite{uhrig2007} to this end. The latter did not perform as well due to its broad spectral content. It remains unknown to us whether the same pulse sequences optimal for dynamic decoupling\cite{gordon2008} are also optimal for the purposes of lock-in measurement.

%%Conclusions
The results presented here demonstrate the potency of the quantum lock-in measurement technique. Since this technique mainly relies on applying time evolution operators that are non-commuting at different times, it is readily available for any quantum probe. Specifically, with single trapped-ions, the lock-in technique enables high precision frequency shift measurements with a nanometer scale spatial resolution. With this method the magnetic field due to the magnetic moment of a, micrometer-separated, co-trapped atomic or molecular ion can be measured. In recent experiments\cite{Koehl2010,Denschlag2010}, trapped ions were submerged inside a quantum degenerate gas of neutral atoms. The quantum lock-in method can be used to probe spin dependent interactions between the ion and the gas, thus serving as local probes with nanometer resolution. Finally, the quantum lock-in technique can be useful for precision measurements and frequency metrology. As an example, it can be used to measure very small frequency shifts required for the observation of parity non-conservation in a single trapped ion\cite{Fortson1993}. Another example is characterizing systematic errors, such as the quadruple shift, in ion-based atomic clocks\cite{Oskay2005}. As a last example, this technique can be used to characterize the noise spectrum of narrow line-width lasers with respect to an atomic transition. We therefore believe that the quantum lock-in technique will be useful for many applications in the field of quantum metrology.

% ************************** METHODS *********************************************************
\begin{methods_summary}
Magnetic noise spectrum is extracted from lock-in contrast measurements by comparison with the theoretical model given by Eq. (\ref{eq:bessel}). To obtain this equation, we assume a magnetic noise Hamiltonian, $H=\frac{g\mu_B}{2}B(t)\hat{\sigma}_z$. The corresponding accumulated lock-in signal is, $\phi_{lock-in}=\frac{g\mu_B}{\hbar}\int_0^TdtB(t)\cos(\int_0^t\Omega(t')dt')$. The lock-in phase scan contrast $A$ is therefore given by averaging over different noise realizations, $(\expect{\cos(\phi_{lock-in})}^2+\expect{\sin(\phi_{lock-in})}^2)^{1/2}$. For a discrete magnetic noise spectrum $B(t)=\sum_n B_n\cos(\omega_n t+\alpha_n)$, and a train of $\pi$ pulses, $\tau_{arm}$ apart, as the lock-in modulation, we obtain Eq. (\ref{eq:bessel}), where averaging is performed over all possible $\{\alpha_n\}$ values.

The low magnetic field noise amplitudes at $50Hz$ and its harmonics reported in this letter are due to an electronic feedback system. We measure the magnetic field using two sense coils at two ends of the vacuum chamber surrounding the ion trap, aligned with the magnetic field quantization axis. The voltages from the two sense coils are fed to pre-amplifiers. A weighted average of these signals, aimed at linearly interpolating the field at the ion position, is used as an error signal of a servo controller which actuates two large compensation coils. This feedback system is able to decrease noise amplitudes by a factor of $10-50$; from approximately $10^{-7}\ T$ to less than $5\times 10^{-9}\ T$. The remaining noise amplitude is due to a non-linear variation of magnetic field noise amplitude and phase across the vacuum chamber. To overcome this non-linearity we installed a feed-forward system, in which the power line voltage is sampled, a phased-locked-loop locks $50,\ 100,\ 150\ Hz$ signals to the line phase, and feeds their sum to the servo controller set-point. We tune the amplitudes and phases of these three synthesized signals, to minimize noise amplitudes at their frequencies, extracted from the lock-in phase scan as described previously. We are thus able to reduce magnetic field noise amplitudes, at the power line frequencies, close to $10^{-10}\ T$.
\end{methods_summary}

% ************************** ADDENDUM *********************************************************
\begin{addendum}
 \item We thank Guy Bensky, Goren Gordon and Gershon Kurizki for helpful discussions.
 \item[Competing Interests] The authors declare that they have no
competing financial interests.
 \item[Correspondence] Correspondence and requests for materials
should be addressed to\\ Shlomi Kotler~ (email: shlomi.kotler@weizmann.ac.il).
\end{addendum}

% ************************** BIBLIOGRAPHY *****************************************************
\bibliography{mybib}

\begin{thebibliography}{10}
\expandafter\ifx\csname url\endcsname\relax
  \def\url#1{\texttt{#1}}\fi
\expandafter\ifx\csname urlprefix\endcsname\relax\def\urlprefix{URL }\fi
\providecommand{\bibinfo}[2]{#2}
\providecommand{\eprint}[2][]{\url{#2}}

\bibitem{Rosenband02008}
\bibinfo{author}{Rosenband, T.} \emph{et~al.}
\newblock \bibinfo{title}{{Frequency Ratio of Al+ and Hg+ Single-Ion Optical
  Clocks; Metrology at the 17th Decimal Place}}.
\newblock \emph{\bibinfo{journal}{Science}} \textbf{\bibinfo{volume}{319}},
  \bibinfo{pages}{1808--1812} (\bibinfo{year}{2008}).

\bibitem{Knunz2010}
\bibinfo{author}{Kn\"unz, S.} \emph{et~al.}
\newblock \bibinfo{title}{Injection locking of a trapped-ion phonon laser}.
\newblock \emph{\bibinfo{journal}{Phys. Rev. Lett.}}
  \textbf{\bibinfo{volume}{105}}, \bibinfo{pages}{013004}
  (\bibinfo{year}{2010}).

\bibitem{Degen2008}
\bibinfo{author}{Degen, C.}
\newblock \bibinfo{title}{{Nanoscale magnetometry microscopy with single
  spins}}.
\newblock \emph{\bibinfo{journal}{{Nature Nanotechnology}}}
  \textbf{\bibinfo{volume}{{3}}}, \bibinfo{pages}{{643--644}}
  (\bibinfo{year}{{2008}}).

\bibitem{Fortson1993}
\bibinfo{author}{Fortson, N.}
\newblock \bibinfo{title}{Possibility of measuring parity nonconservation with
  a single trapped atomic ion}.
\newblock \emph{\bibinfo{journal}{Phys. Rev. Lett.}}
  \textbf{\bibinfo{volume}{70}}, \bibinfo{pages}{2383--2386}
  (\bibinfo{year}{1993}).

\bibitem{Kielpinski2001}
\bibinfo{author}{Kielpinski, D.} \emph{et~al.}
\newblock \bibinfo{title}{A decoherence-free quantum memory using trapped
  ions}.
\newblock \emph{\bibinfo{journal}{Science}} \textbf{\bibinfo{volume}{291}},
  \bibinfo{pages}{1013--1015} (\bibinfo{year}{2001}).

\bibitem{Roos2004}
\bibinfo{author}{Roos, C.~F.} \emph{et~al.}
\newblock \bibinfo{title}{Bell states of atoms with ultralong lifetimes and
  their tomographic state analysis}.
\newblock \emph{\bibinfo{journal}{Phys. Rev. Lett.}}
  \textbf{\bibinfo{volume}{92}}, \bibinfo{pages}{220402}
  (\bibinfo{year}{2004}).

\bibitem{Langer2005}
\bibinfo{author}{Langer, C.} \emph{et~al.}
\newblock \bibinfo{title}{Long-lived qubit memory using atomic ions}.
\newblock \emph{\bibinfo{journal}{Phys. Rev. Lett.}}
  \textbf{\bibinfo{volume}{95}}, \bibinfo{pages}{060502}
  (\bibinfo{year}{2005}).

\bibitem{Leibfried2004}
\bibinfo{author}{Leibfried, D.} \emph{et~al.}
\newblock \bibinfo{title}{Toward heisenberg-limited spectroscopy with
  multiparticle entangled states}.
\newblock \emph{\bibinfo{journal}{Science}} \textbf{\bibinfo{volume}{304}},
  \bibinfo{pages}{1476--1478} (\bibinfo{year}{2004}).

\bibitem{Roos2006}
\bibinfo{author}{Roos, C.~F.}, \bibinfo{author}{Chwalla, M.},
  \bibinfo{author}{Kim, K.}, \bibinfo{author}{Riebe, M.} \&
  \bibinfo{author}{Blatt, R.}
\newblock \bibinfo{title}{`designer atoms' for quantum metrology}.
\newblock \emph{\bibinfo{journal}{Nature}} \textbf{\bibinfo{volume}{443}},
  \bibinfo{pages}{316--319} (\bibinfo{year}{2006}).

\bibitem{Cirac1997}
\bibinfo{author}{Huelga, S.~F.} \emph{et~al.}
\newblock \bibinfo{title}{Improvement of frequency standards with quantum
  entanglement}.
\newblock \emph{\bibinfo{journal}{Phys. Rev. Lett.}}
  \textbf{\bibinfo{volume}{79}}, \bibinfo{pages}{3865--3868}
  (\bibinfo{year}{1997}).

\bibitem{Lukin2004}
\bibinfo{author}{Andr\'e, A.}, \bibinfo{author}{S\o{}rensen, A.~S.} \&
  \bibinfo{author}{Lukin, M.~D.}
\newblock \bibinfo{title}{Stability of atomic clocks based on entangled atoms}.
\newblock \emph{\bibinfo{journal}{Phys. Rev. Lett.}}
  \textbf{\bibinfo{volume}{92}}, \bibinfo{pages}{230801}
  (\bibinfo{year}{2004}).

\bibitem{Uys2009}
\bibinfo{author}{Uys, H.}, \bibinfo{author}{Biercuk, M.~J.} \&
  \bibinfo{author}{Bollinger, J.~J.}
\newblock \bibinfo{title}{Optimized noise filtration through dynamical
  decoupling}.
\newblock \emph{\bibinfo{journal}{Physical Review Letters}}
  \textbf{\bibinfo{volume}{103}} (\bibinfo{year}{2009}).

\bibitem{Biercuk2009}
\bibinfo{author}{Biercuk, M.~J.} \emph{et~al.}
\newblock \bibinfo{title}{Optimized dynamical decoupling in a model quantum
  memory}.
\newblock \emph{\bibinfo{journal}{Nature}} \textbf{\bibinfo{volume}{458}},
  \bibinfo{pages}{996--1000} (\bibinfo{year}{2009}).

\bibitem{goren:2007}
\bibinfo{author}{Gordon, G.}, \bibinfo{author}{Erez, N.} \&
  \bibinfo{author}{Kurizki, G.}
\newblock \bibinfo{title}{Universal dynamical decoherence control of noisy
  single- and multi-qubit systems}.
\newblock \emph{\bibinfo{journal}{Journal of Physics B: Atomic, Molecular and
  Optical Physics}} \textbf{\bibinfo{volume}{40}}, \bibinfo{pages}{S75--S93}
  (\bibinfo{year}{2007}).

\bibitem{Maze2008}
\bibinfo{author}{Maze, J.~R.} \emph{et~al.}
\newblock \bibinfo{title}{{Nanoscale magnetic sensing with an individual
  electronic spin in diamond}}.
\newblock \emph{\bibinfo{journal}{{Nature}}} \textbf{\bibinfo{volume}{{455}}},
  \bibinfo{pages}{{644--648}} (\bibinfo{year}{{2008}}).

\bibitem{Hall2010}
\bibinfo{author}{Hall, L.~T.}, \bibinfo{author}{Hill, C.~D.},
  \bibinfo{author}{Cole, J.~H.} \& \bibinfo{author}{Hollenberg, L. C.~L.}
\newblock \bibinfo{title}{Ultrasensitive diamond magnetometry using optimal
  dynamic decoupling}.
\newblock \emph{\bibinfo{journal}{Phys. Rev. B}} \textbf{\bibinfo{volume}{82}},
  \bibinfo{pages}{045208} (\bibinfo{year}{2010}).

\bibitem{Naydenov2010}
\bibinfo{author}{{Naydenov}, B.} \emph{et~al.}
\newblock \bibinfo{title}{{Dynamical Decoupling of a single electron spin at
  room temperature}}.
\newblock \emph{\bibinfo{journal}{ArXiv e-prints}}  (\bibinfo{year}{2010}).
\newblock \eprint{1008.1953}.

\bibitem{deLange2010_2}
\bibinfo{author}{{de Lange}, G.}, \bibinfo{author}{{Rist{\`e}}, D.},
  \bibinfo{author}{{Dobrovitski}, V.~V.} \& \bibinfo{author}{{Hanson}, R.}
\newblock \bibinfo{title}{{Single-spin magnetometry with multi-pulse dynamical
  decoupling sequences}}.
\newblock \emph{\bibinfo{journal}{ArXiv e-prints}}  (\bibinfo{year}{2010}).
\newblock \eprint{1008.4395}.

\bibitem{Balasubramanian2009}
\bibinfo{author}{Balasubramanian, G.} \emph{et~al.}
\newblock \bibinfo{title}{{Ultralong spin coherence time in isotopically
  engineered diamond}}.
\newblock \emph{\bibinfo{journal}{{Nature Materials}}}
  \textbf{\bibinfo{volume}{{8}}}, \bibinfo{pages}{{383--387}}
  (\bibinfo{year}{{2009}}).

\bibitem{Kesselman2010}
\bibinfo{author}{Kesselman, A.}, \bibinfo{author}{Glickman, Y.},
  \bibinfo{author}{Akerman, N.}, \bibinfo{author}{Kotler, S.} \&
  \bibinfo{author}{Ozeri, R.}
\newblock \bibinfo{title}{In preparation} .

\bibitem{GillMeiboom1958}
\bibinfo{author}{Meiboom, S.} \& \bibinfo{author}{Gill, D.}
\newblock \bibinfo{title}{Modified spin-echo method for measuring nuclear
  relaxation times}.
\newblock \emph{\bibinfo{journal}{Review of Scientific Instruments}}
  \textbf{\bibinfo{volume}{29}}, \bibinfo{pages}{688--691}
  (\bibinfo{year}{1958}).

\bibitem{preprint}
\bibinfo{author}{Kotler, S.}, \bibinfo{author}{Akerman, N.},
  \bibinfo{author}{Glickman, Y.} \& \bibinfo{author}{Ozeri, R.}
\newblock \bibinfo{title}{In preparation} .

\bibitem{AVAR}
\bibinfo{author}{Allan, D.~W.}
\newblock \bibinfo{title}{Allan variance}.
\newblock \urlprefix\url{http://www.allanstime.com}.

\bibitem{PDD}
\bibinfo{author}{Drever, R. W.~P.} \emph{et~al.}
\newblock \bibinfo{title}{Laser phase and frequency stabilization using an
  optical resonator}.
\newblock \emph{\bibinfo{journal}{Applied Physics B: Lasers and Optics}}
  \textbf{\bibinfo{volume}{31}}, \bibinfo{pages}{97--105}
  (\bibinfo{year}{1983}).

\bibitem{uhrig2007}
\bibinfo{author}{Uhrig, G.~S.}
\newblock \bibinfo{title}{Keeping a quantum bit alive by optimized
  $\pi{}$-pulse sequences}.
\newblock \emph{\bibinfo{journal}{Phys. Rev. Lett.}}
  \textbf{\bibinfo{volume}{98}}, \bibinfo{pages}{100504}
  (\bibinfo{year}{2007}).

\bibitem{gordon2008}
\bibinfo{author}{Gordon, G.}, \bibinfo{author}{Kurizki, G.} \&
  \bibinfo{author}{Lidar, D.~A.}
\newblock \bibinfo{title}{Optimal dynamical decoherence control of a qubit}.
\newblock \emph{\bibinfo{journal}{Phys. Rev. Lett.}}
  \textbf{\bibinfo{volume}{101}}, \bibinfo{pages}{010403}
  (\bibinfo{year}{2008}).

\bibitem{Koehl2010}
\bibinfo{author}{Zipkes, C.}, \bibinfo{author}{Palzer, S.},
  \bibinfo{author}{Sias, C.} \& \bibinfo{author}{Koehl, M.}
\newblock \bibinfo{title}{A trapped single ion inside a bose-einstein
  condensate}.
\newblock \emph{\bibinfo{journal}{Nature}} \textbf{\bibinfo{volume}{464}},
  \bibinfo{pages}{388--391} (\bibinfo{year}{2010}).

\bibitem{Denschlag2010}
\bibinfo{author}{Schmid, S.}, \bibinfo{author}{H\"arter, A.} \&
  \bibinfo{author}{Denschlag, J.~H.}
\newblock \bibinfo{title}{Dynamics of a cold trapped ion in a bose-einstein
  condensate}.
\newblock \emph{\bibinfo{journal}{Phys. Rev. Lett.}}
  \textbf{\bibinfo{volume}{105}}, \bibinfo{pages}{133202}
  (\bibinfo{year}{2010}).

\bibitem{Oskay2005}
\bibinfo{author}{Oskay, W.~H.}, \bibinfo{author}{Itano, W.~M.} \&
  \bibinfo{author}{Bergquist, J.~C.}
\newblock \bibinfo{title}{Measurement of the $^{199}hg^{+}$
  $5d^{9}6s^{2}\,^{2}d_{5/2}$ electric quadrupole moment and a constraint on
  the quadrupole shift}.
\newblock \emph{\bibinfo{journal}{Phys. Rev. Lett.}}
  \textbf{\bibinfo{volume}{94}}, \bibinfo{pages}{163001}
  (\bibinfo{year}{2005}).

\end{thebibliography}
\bibliographystyle{naturemag}

		\includegraphics[scale=0.95]{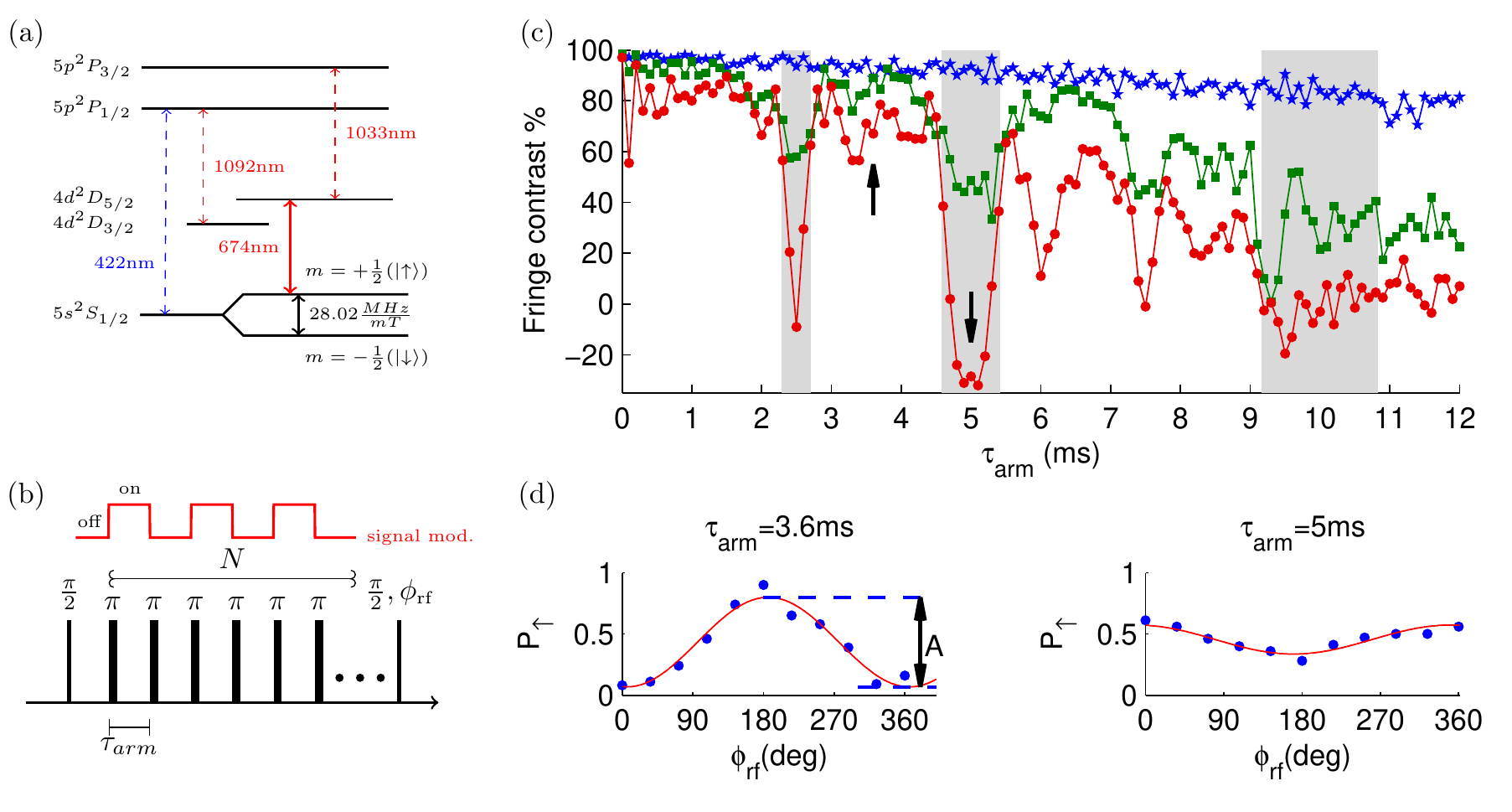}
\begin{figure} % ************** Multiple Echo figure ******************************
		
    \caption{
    Measurement scheme.
    \textbf{(a)} $\ ^{88}Sr^+$ level diagram. The probe spin states are $\up=|5s_{1/2}, m=+1/2\rangle$ and $\down=|5s_{1/2}, m=-1/2\rangle$. An external magnetic field of $204\ \mu T$ splits the two levels by a frequency of $\omega_0=(2\pi)5.72\ MHz$ and determines a quantization axis. Spin rotations are performed using an oscillating magnetic field which is perpendicular to the quantization axis. Spin detection is performed by shelving the $\up$ state to the meta-stable $|4d_{5/2},m=+3/2\rangle$ level, with a narrow line-width ($<100\ Hz$) $674\ nm$ laser, followed by detecting state selective fluorescence on the $422nm$ transition. Lasers at $1092nm$ and $1033nm$ are used for repumping the meta-stable $D$ levels.
    \textbf{(b)} The quantum lock-in measurement pulse scheme. The ion is initialized to $\frac{1}{\sqrt{2}}(\up + \down)$. While the measured signal is modulated, the superposition is also modulated, in phase with the signal, by a train of $N$ $\pi$-pulses, $\tau_{arm}$ apart. The total relative phase, $\phi_{lock-in}$, of the ion superposition $\frac{1}{\sqrt{2}}(\down+e^{i\phi_{lock-in}}\up)$ accumulated during the lock-in sequence is measured by scanning the phase of a final $\pi/2$ pulse, $\phi_{rf}$ followed by spin detection.
    \textbf{(c)} Phase scan contrast vs. lock-in modulation period $2\tau_{arm}$, in the absence of any modulated signal. Plots corresponding to $N=1,9$ and $17$-$\pi$ pulses are shown in blue stars, green rectangles and red circles respectively. We observe contrast drops as $\tau_{arm}$ approaches $2.5\ ms,5\ ms$ and $10\ ms$ corresponding to $200Hz,100Hz$ and $50Hz$ magnetic field noise components.
    \textbf{(d)} Probability of finding the ion in the $\up$ state vs. $\phi_{rf}$. Phase scan plots for $\tau_{arm}=3.6\ ms$ and $\tau_{arm}=5\ ms$ are shown. Both are taken with an $N=17$ $\pi$-pulses lock-in sequence. Each point is the average of $100$ experiments. Solid line is a best-fit to $P(\uparrow) =\frac{1}{2}+\frac{A}{2}\cos(\phi_{lock-in}+\phi_{rf})$. The fitted $A$ are shown in (c) at the locations indicated by the two black arrows.
    }
\end{figure}

\begin{figure} %************** Sensitiviy Figure ********************************
	\includegraphics[scale=0.95]{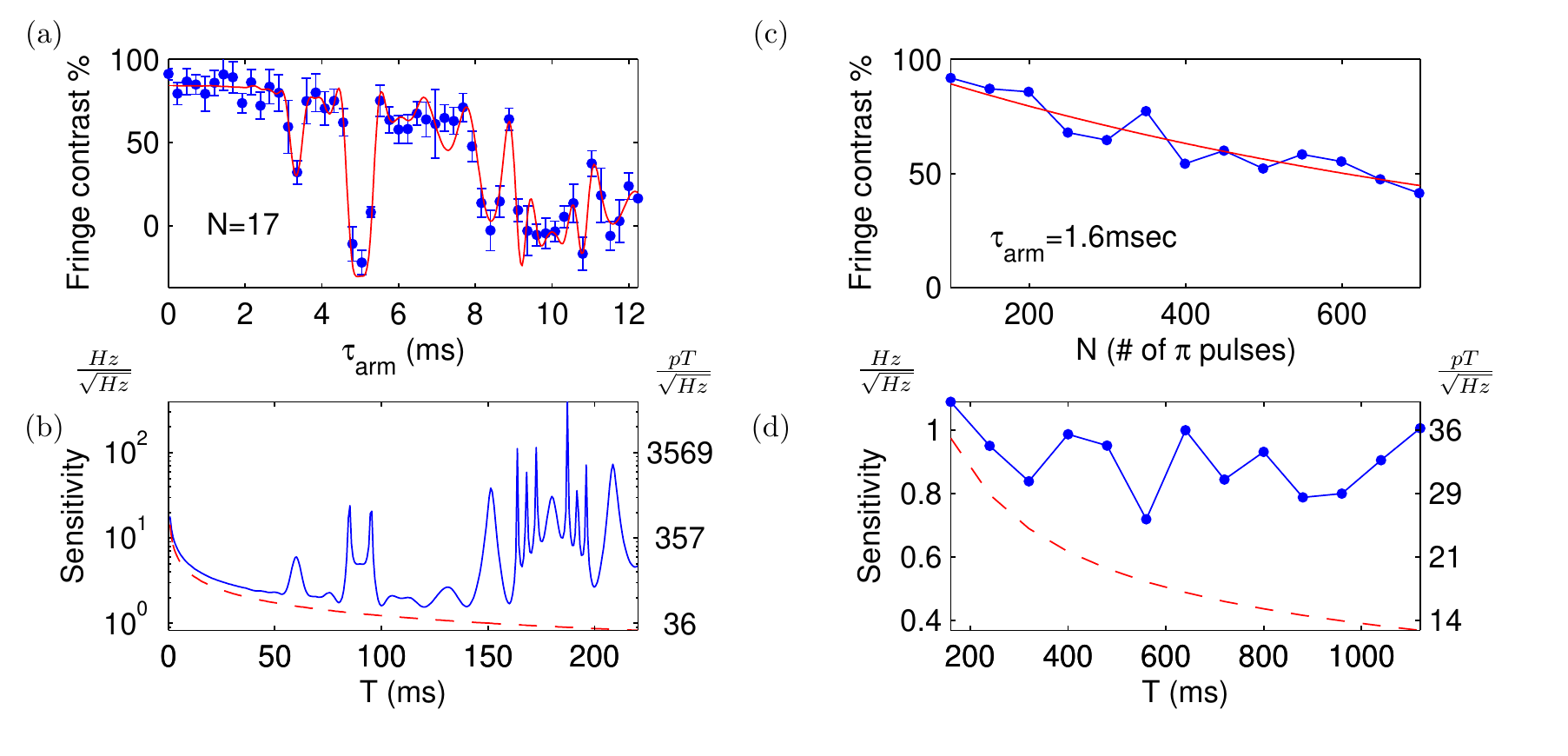}
	\caption{
	Noise floor and sensitivity of the quantum lock-in measurement.
	\textbf{(a)} Phase scan contrast vs. $\tau_{arm}$ for $N=17$ $\pi$-pulses, in the absence of any modulated signal. We extract the magnetic noise spectrum by using Eq. (\ref{eq:bessel}). Solid red line is a best fit to Eq. (\ref{eq:bessel}) where fit parameters are the field amplitudes $B_{50 Hz}=540(3)\ pT,\ B_{100 Hz}=390(5)\ pT,\ B_{150 Hz}=260(4)\ pT$ and a slowly varying magnetic field $g\mu_{B}B_\textrm{slow}f_\textrm{slow}/h=37(4)Hz^2$.
	\textbf{(b)} Solid blue line is the lock-in sensitivity vs. the total lock-in sequence time, calculated from (a) using Eq. (\ref{eq:phSensitivity}). The dashed red line is the standard quantum limit to sensitivity (achieved when $A=1$). A best sensitivity of $1.53\ \frac{Hz}{\sqrt{Hz}}$ ($53\ \frac{pT}{\sqrt{Hz}}$) is observed at a lock-in sequence duration of $120\ ms$. This sensitivity is only a factor of 1.5 larger than the standard quantum limit.
	\textbf{(c)} Phase scan contrast vs. number of $\pi$-pulses, $N$, at a lock-in modulation period $2\tau_{arm}=3.2\ ms$. The red line is an exponential decay fit to the data yielding a 1/e coherence decay time of $1.4(2)\ s$.
	\textbf{(d)} As in (b), the blue solid line is lock-in sensitivity calculated from (c). Dashed red line shows the standard quantum limit to the lock-in sensitivity. A best sensitivity of $0.72\ \frac{Hz}{\sqrt{Hz}}$ ($25.5\ \frac{pT}{\sqrt{Hz}}$) is observed at a lock-in sequence duration of $T=560\ ms$.
	}
\end{figure}

\begin{figure} % ********************* Lock-In Detection Figure ******************************************
	\includegraphics{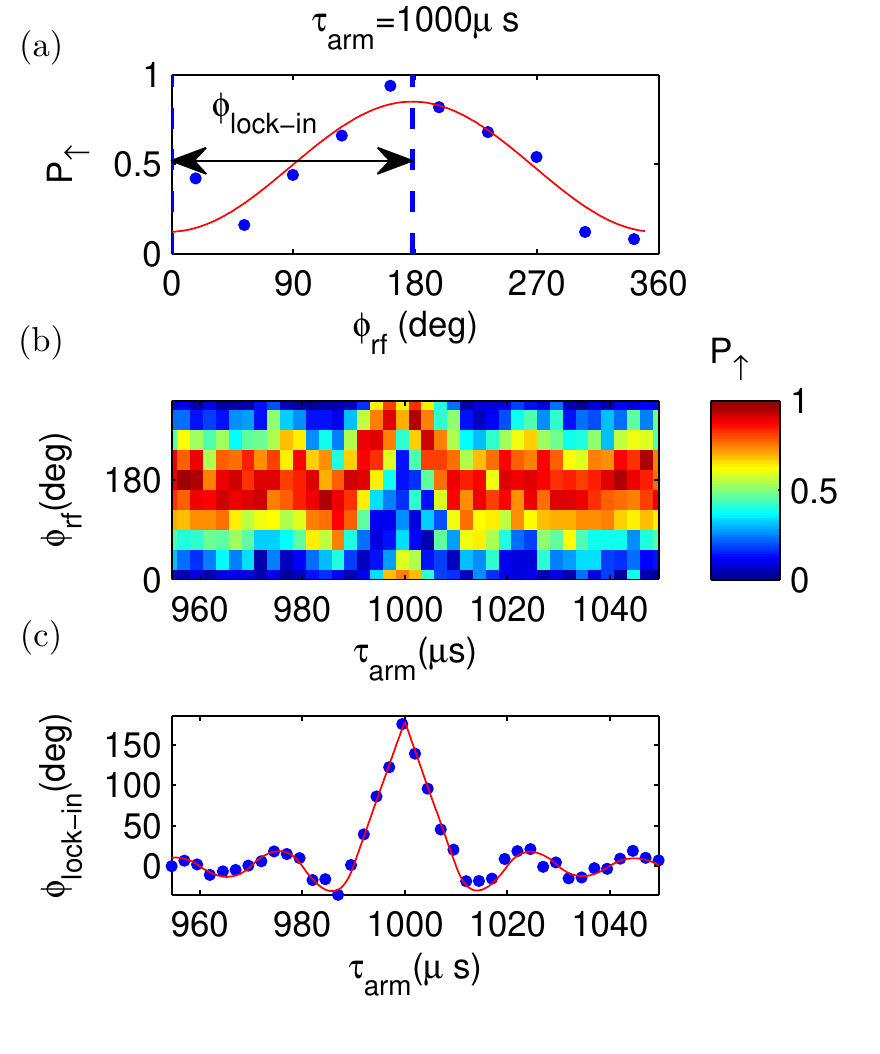}
	\caption{
	Lock-in measurement of a small signal. The light shift of the $\up$ state, induced by the $674nm$ laser is measured. The laser is detuned by $\delta_{674nm}=-17kHz$ from the $\up \rightarrow |4d_{5/2}, m=+3/2\rangle$ quadruple transition, and is amplitude modulated by a square wave of frequency $f_L=500Hz$. The lock-in scheme has $N=99$ $\pi$-pulses while the lock-in frequency $f_m=1/2\tau_{arm}$ is varied.
	\textbf{(a)} Lock-in phase scan; $P_{\uparrow}$ vs. $\phi_{rf}$, at a lock-in period $2\tau_{arm}=2\ ms$. Red solid line is a best fit to $P_{\uparrow}=\frac{1}{2}+\frac{A}{2}\cos(\phi_{lock-in}+\phi_{rf})$. A clear phase shift of $\phi_{lock-in}=0.99\pi\ rad$ is observed.
	\textbf{(b)} Every column is a lock-in phase scan similar to (a) for various values of $\tau_{arm}$. The lock-in signal, $\phi_{lock-in}$ is seen to increase as the lock-in modulation frequency $f_m$ approaches the laser modulation frequency $f_L=500Hz$.
	\textbf{(c)} The fitted $\phi_{lock-in}$ vs. $\tau_{arm}$, extracted from (b) as explained in (a). A light shift of $9.7(4)\ Hz$ is measured (with $95\%$ confidence). The solid red line is calculated using Eq. (\ref{eq:accPhase}) without any fit parameters.
	}
\end{figure}

\begin{figure} %*********************** Light Shift Figure ***********************************************
	\includegraphics{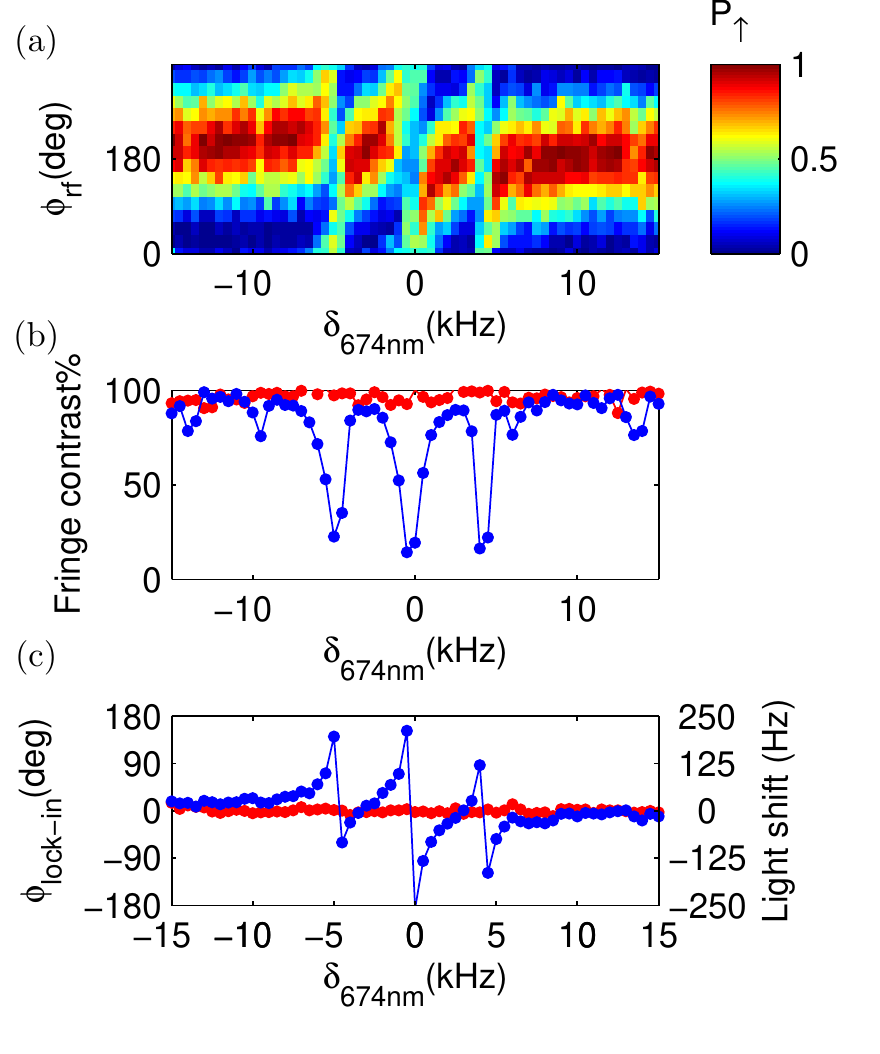}
	\caption{
	Light shift spectroscopy. Light shift of $\up$ induced by the $674nm$ laser vs. laser detuning $\delta_{674nm}$. At each given $\delta_{674nm}$ a lock-in sequence of $N=39$ $\pi$-pulses with a lock-in period of $2\tau_{arm}=200\mu s$ is applied while the $674nm$ laser is amplitude modulated at the same frequency.
    \textbf{(a)} Every column is a lock-in phase scan for various values of $\delta_{674nm}$.
	\textbf{(b)} Blue filled circles are the fitted phase scan contrast, $A$, vs. $\delta_{674nm}$.  Red filled circles are the fitted $A$ in the absence of the $674nm$ laser light. A reduction in contrast is observed due to shelving of the $\up$ state to the metastable D level whenever the laser approaches resonance.
	\textbf{(c)} Blue filled circles are the fitted $\phi_{lock-in}$ vs. $\delta_{674nm}$. Red filled circles are $\delta_{lock-in}$ in the absence of the 674nm laser light. Light shifts are seen to have dispersive resonance. Both (b) and (c) show two sidebands, separated by $5\ kHz$ from the transition carrier, generated by the fast amplitude modulation of the laser.
	}
\end{figure}

\end{document}